\documentclass[pra, superscriptaddress, showpacs]{revtex4}
\usepackage{graphicx}

\begin{document}
\title{Experimentally realizable quantum comparison of coherent states and its applications}
\author{Erika Andersson}
\affiliation{Department of Physics, SUPA, University of Strathclyde, Glasgow G4 0NG, UK}
\author{Marcos Curty}
\affiliation{Quantum Information Theory Group, Institut f\"ur
Theoretische Physik I, and Max-Planck Research Group, Institute
of Optics, Information and Photonics, Universit\"{a}t
Erlangen-N\"{u}rnberg, Staudtstra{\ss}e 7/B2, 91058 Erlangen,
Germany}
%\footnote{Corresponding author: email erika@phys.strath.ac.uk}
\author{Igor Jex}
\affiliation{Department of Physics, FNSPE Czech Technical University in Prague\\ B\v rehov\'a 7, 115 19 Praha 1, Czech Republic}
\date{\today}

\begin{abstract}
When comparing quantum states to each other, it is possible to obtain an unambiguous answer, indicating that the states are definitely different, already after a single measurement. In this
paper we investigate comparison of coherent states, which is the simplest example of quantum state
comparison for continuous variables. 
The method we present has a high success probability, and  is experimentally feasible to realize as the only required components are beam splitters and photon detectors.
An easily realizable method for quantum state comparison could be important for real applications. As examples of such applications we present a ``lock and key" scheme and a simple scheme for quantum
public key distribution.
\end{abstract}
%Quantum information, quantum cryptography, quantum optics, Foundations of quantum mechanics; measurement theory 03.67.Hk -- quantum communication
\maketitle
%{\it Keywords:} Quantum information; Quantum comparison; Quantum cryptography; Linear optics; Interferometry
%\\
%\\
%Igor Jex\\
%phone: +420 22435 8264\\
%fax: +420 222 320 861\\
%e-mail: igor.jex@fjfi.cvut.cz\\
%\\
%Erika Andersson\\
%phone: +44 141 548 3376\\
%fax: +44 141 552 2891\\
%e-mail: erika@phys.strath.ac.uk

\section{Introduction}

The generation, manipulation and measurement of individual quantum objects has become everyday practice in the laboratory. Many experiments have proven that we have the technological means to
perform a wide range of manipulations, which, just after the advent of quantum mechanics, one could only dream of.

Because of fundamental differences between classical and quantum objects, certain operations cannot be performed in the quantum domain at all, or they can be performed only with a fidelity less than one. The list of such operations is already quite long, including for instance cloning \cite{zurekwootters} and entangling. Neither of these processes can be performed perfectly unless we know the initial state (or which set of orthogonal states the initial state belongs to).
Such operations are {\it universal}, in the sense that we are aiming at performing a transformation or manipulation of the quantum object, independent of the exact form of the input. 
Another example of such a universal process would be state comparison. We can ask whether two given (pure) quantum states are identical or not. If no a priori information about the states is available, we have to limit ourselves to looking at the inherent symmetry of our two particle system with respect to permutation. The total state for two identical quantum states is always symmetric, and therefore asymmetry is the unambiguous indicator of dissimilarity.
 
Comparison of unknown as well as completely known quantum states has been analyzed in detail \cite{firstcomp, statecomp, tonycomp, dagmarcomp}, as well as comparison of unitary transforms \cite{trafocomp}.  Not much attention, however, was paid to cases where partial knowledge about the possible  states on which the comparison should be performed is available. In the present paper we wish to concentrate on this particular case. We choose to look at comparison of coherent states. The unknown parameter in the states to be compared is the coherent state amplitude $\alpha$, specified by two numbers -- its absolute value and its phase \cite{Glauber}. The reason for choosing coherent states is that they are easy to generate and convenient to use. Another aspect is that present suggestions for realizing quantum comparison either require non-trivial components (CNOT gates as in the swap test \cite{finger, digital}) or destroy the states to be compared (multiport implementation of universal comparison \cite{statecomp}). Coherent states, on the other hand, may be compared non-invasively, using only linear optics and photon detectors, if they are identical (meaning that they are equal both in phase and in amplitude). 
In the following we discuss not only the question of comparing two or more coherent states to each other, but we also analyze two possible applications. We present a simple ``lock and key" scheme and a public key distribution scheme using coherent state comparison as an essential ingredient.

\section{Comparison of coherent states}

\subsection{Two coherent states}

Let us first see how to determine whether two Glauber coherent states $|\alpha\rangle$ and $|\beta\rangle$ \cite{Glauber} are different from each other. A coherent state is a state for which $\hat{a}|\alpha\rangle = \alpha|\alpha\rangle$, where $\hat{a}$ is the annihilation operator for the concerned electromagnetic field mode. Here we have no knowledge of the amplitude or phase of $\alpha$ and $\beta$, but we do know that the states are coherent.
The two states can be compared using a 50/50 beam splitter in the following way.
\begin{figure}
\includegraphics[width=5.cm]{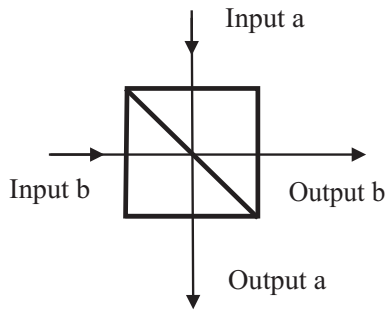}
\caption{The beam splitter mixes the two input fields in a linear
way into two output fields. The relations between input and output creation operators are expressed in Eq. (1).} \label{bs1}
\end{figure}
It is well known, that if two coherent states $|\alpha\rangle$ and
$|\beta\rangle$ are incident on a balanced beam splitter, as shown in Fig. \ref{bs1}, then the
output states will be $|(\alpha+\beta)/\sqrt{2}\rangle$
and $|(\alpha-\beta)/\sqrt{2}\rangle$ \cite{loudon}. This follows since the beam splitter relations between the input 
and output creation operators  are
\begin{eqnarray}
\hat{a}^\dagger_{out} &=& \frac{1}{\sqrt{2}}(\hat{a}^\dagger_{in}+\hat{b}^\dagger_{in})\nonumber\\
\hat{b}^\dagger_{out} &=& \frac{1}{\sqrt{2}}(\hat{a}^\dagger_{in}-\hat{b}^\dagger_{in}).
\end{eqnarray}
Using these relations, we can confirm that the coherent states $|\alpha\rangle$ and $|\beta\rangle$ transform as
\begin{eqnarray}
|\alpha\rangle_{a,in}\otimes|\beta\rangle_{b,in}
&=& e^{-\frac{1}{2}|\alpha|^2} \exp{(\alpha \hat{a}^\dagger_{in})}
 e^{-\frac{1}{2}|\beta|^2} \exp(\beta \hat{b}^\dagger_{in})|0\rangle\nonumber \\
&=& e^{-\frac{1}{2}(|\alpha|^2+|\beta|^2)} \exp\left[\frac{\alpha}{\sqrt{2}}(\hat{a}^\dagger_{out}+\hat{b}^\dagger_{out})+\frac{\beta}{\sqrt{2}}(\hat{a}^\dagger_{out}-\hat{b}^\dagger_{out})\right]|0\rangle\nonumber\\
&=& e^{-\frac{1}{2}(|\alpha|^2+|\beta|^2)} \exp\left[\frac{1}{\sqrt{2}}(\alpha+\beta)\hat{a}^\dagger_{out}+\frac{1}{\sqrt{2}}(\alpha-\beta)\hat{b}^\dagger_{out}\right]|0\rangle\nonumber\\
&=&|\frac{\alpha+\beta}{\sqrt{2}}\rangle_{a,out}\otimes|\frac{\alpha-\beta}{\sqrt{2}}\rangle_{b,out}.
\end{eqnarray}
If $\alpha$ and $\beta$ are equal, output mode $b$ will contain only the vacuum. Therefore, if we detect any number of photons in this mode, we can be certain that $\alpha$ and $\beta$ cannot have been identical both in phase and amplitude. We have of course assumed that there are no dark counts in
the detectors. If the probability for dark counts is non-zero, we cannot anymore infer with certainty that $\alpha$ and $\beta$ were different.
Detector inefficiency is not as crucial as dark counts. An efficiency less than one will of course degrade the probability of detecting a difference, but will not prevent us from drawing the conclusion that $\alpha$ and $\beta$ {\it must} have been different. This is because each detector click in output mode $b$, which is not a dark count, is a valid indicator of difference between the input states. If some of photons in output mode $b$ are not detected, this will decrease the efficiency of difference detection, but does not make it impossible to infer that $\alpha\neq\beta$.

\begin{figure}
\includegraphics[width=8.cm]{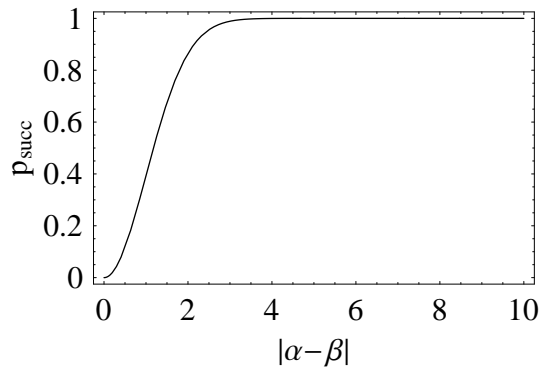}
\caption{The success probability of comparing two coherent states as a function of the absolute value of the difference between the coherent states amplitudes, $\vert\alpha -\beta\vert$. The knowledge that we deal with an a priori known class of states enables us to reach the ideal limit of $1$.} \label{Fig1cp}
\end{figure}

The success probability of detecting a difference between $\alpha$ and $\beta$ is equal to the probability to detect at least one photon in output mode $b$, where we have the coherent state $|(\alpha-\beta)/\sqrt{2}\rangle$. As the probability to detect zero photons in this mode
is $p(0) = \exp(-1/2|\alpha-\beta|^2)$, the success probability is
\begin{equation}
p_{succ} = 1-p(0) = 1 - e^{-\frac{1}{2}|\alpha-\beta|^2}. \label{psucctwo}
\end{equation}
The success probability increases exponentially to its maximum value of $1$ as shown in Fig. \ref{Fig1cp}.

A nice feature of this method is that we do not need to place any detector in output mode $a$. This means that we can again split the state in output mode $a$, $|(\alpha+\beta)/\sqrt{2}\rangle$,
with a second 50/50 beam splitter, giving the output $|(\alpha+\beta)/2\rangle|(\alpha+\beta)/2\rangle$. If no photons were found in mode $b$, and $\alpha$ and $\beta$ were indeed equal both in phase and amplitude, we recover the original states undisturbed. The fact that the states emerge undisturbed indicates a non-demolition aspect of the state comparison procedure, which could
be useful for applications where quantum state comparison is needed. It should be pointed out, however, that if no photons are detected in mode $b$, we cannot actually be sure that the coherent
states were really identical. If they are not, and the output state is again split by the second beam splitter, the resulting states will differ from $|\alpha\rangle$ and $|\beta\rangle$. They will %be averaged, and 
both be equal to $|(\alpha+\beta)/2\rangle$.

In a similar way, if we choose to detect photons in output mode $a$ instead of $b$, we can conclude that $\alpha$ and $-\beta$ cannot have been identical. The success probability for this is $p'_{succ} = 1-\exp(-|\alpha+\beta|^2/2)$. No matter in which output we detect one or more photons, a detector click will give us information about the input states. If we know, for instance, that $|\alpha|=|\beta|$, the detector clicks will give information about the relative phase $\delta$ of $\alpha$ and $\beta$. Using the just described method, we obtain the success probability 
\begin{equation}
p_{succ} = 1-p(0) = 1 - e^{-\vert\alpha\vert^2\sin^2(\delta/2)}
\end{equation}
This result indicates, that if the phase difference is large, moderate coherent amplitudes are already sufficient to yield a high success probability for the comparison test. For small phase differences, this is possible only for large amplitudes. To obtain a sufficient success probability, the phase difference should scale according to 
$\delta\approx c/\alpha$, where $c$ is a constant, and $\alpha$ is the coherent state amplitude. Hence the comparison strategy does not offer any particular advantage when searching for optimal phase measurements \cite{phase}.

Finally, we could instead use a beam splitter which is not balanced, but has different transmission and reflection coefficients $T$ and $R$. In this case, the output state is given by $|\sqrt{T}\alpha+\sqrt{R}\beta\rangle\otimes|\sqrt{R}\alpha-\sqrt{T}\beta\rangle$.
Finding photons in the first output mode determines that $\sqrt{T}\alpha+\sqrt{R}\beta\neq 0$, and photons in the second output mode means that $\sqrt{R}\alpha-\sqrt{T}\beta \neq 0$. With
phase shifters before the input ports of the beam splitter, we can more generally test whether $\sqrt{T} e^{i\theta} \alpha+\sqrt{R}\beta\neq 0$.

We can also compare the success probability (\ref{psucctwo}) with the success probability for the universal comparison strategy. If we want to compare two general pure quantum states $|\phi\rangle$ and $|\psi\rangle$, but we have no information about the states, the best we can do is to check whether the overall state $|\phi\rangle\otimes|\psi\rangle$ is symmetric with respect to permutation or not  \cite{firstcomp, statecomp}. If $|\phi\rangle$ and $|\psi\rangle$ are equal, the overall state is necessarily
symmetric. Therefore, if the state is found not to be symmetric, we can be sure that the states were not equal. The success probability is the probability of finding the states in the asymmetric subspace, which is
\begin{equation}
\label{psuccuniv}
p_{asymm} =1-p_{symm} = \frac{1}{2}(1-|\langle\phi|\psi\rangle|^2).
\end{equation}
For the two coherent states, this success probability becomes
\begin{equation}
p_{asymm} = \frac{1}{2}(1-e^{-|\alpha-\beta|^2}).
\end{equation}
The success probability (\ref{psucctwo}) for the coherent state comparison is larger than that of the optimal universal comparison strategy, since
\begin{eqnarray}
(1-e^{-\frac{1}{2}|\alpha-\beta|^2})^2 &\geq & 0 \iff \nonumber\\
p_{succ} = 1 - e^{-\frac{1}{2}|\alpha-\beta|^2} &\geq & \frac{1}{2}(1-e^{-|\alpha-\beta|^2}) = p_{asymm},
\end{eqnarray}
with equality only when both probabilities are zero, i.e. when $\alpha = \beta$. We are able to obtain a ``better than optimal" success probability since, in the above beam splitter scheme, we made use of the additional knowledge that the states are coherent. If we would not have this knowledge, we would have to revert to the universal comparison strategy. The success probability of the optimal universal strategy is always below 1/2, whereas when $|\alpha-\beta|$ becomes large,  the success probability of the beam
splitter strategy approaches one. This reflects the fact, that when $|\alpha-\beta|$
is large, we enter the classical regime. Here we have not addressed the question whether the beam splitter strategy is optimal for coherent states. It certainly has the appealing feature that it is easy to
implement experimentally, which is very important. 

\subsection{Comparing squeezed vacua}

Squeezed vacua may also be compared to each other using a beam splitter. A beam splitter transforms two squeezed
vacua $s_1\exp(\xi_1\hat{a}^{\dagger 2}_{in})|0\rangle$ and $s_2\exp(\xi_2\hat{b}^{\dagger 2}_{in})|0\rangle$, where $s_1$ and $s_2$ are normalization constants,
according to
\begin{eqnarray}
s_1 s_2 \exp(\xi_1\hat{a}^{\dagger 2}_{in} +\xi_2\hat{b}^{\dagger 2}_{in})|0\rangle &=&
s_1 s_2 \exp\{\frac{1}{2}[\xi_1(\hat{a}^\dagger_{out}+\hat{b}^\dagger_{out})^2+\xi_2(\hat{a}^\dagger_{out}-\hat{b}^\dagger_{out})^2]\}|0\rangle\\
&=&s_1 s_2 \exp[\frac{1}{2}(\xi_1+\xi_2)(\hat{a}^{\dagger 2}_{out}+\hat{b}^{\dagger 2}_{out})+ (\xi_1-\xi_2)\hat{a}^\dagger_{out}\hat{b}^\dagger_{out}]|0\rangle.\nonumber
\end{eqnarray}
From this expression, we see that when $\xi_1=\xi_2$, both output modes will contain only even numbers of photons. Detecting an odd number of photons in either of the outputs (assuming perfect
detectors) therefore indicates that $\xi_1\neq\xi_2$. Correspondingly, detecting an even number of photons indicates that $\xi_1\neq-\xi_2$. The expression for the probability to detect an odd number of photon is rather cumbersome. It takes the explicit form
\begin{eqnarray}
p_{2l +1, 2m+1} &=&
s_1 s_2\sum_i \sum\limits^{min \{l,m\}}_{k=0} \frac{(\xi_1
-\xi_2)^k}{k!} \sqrt{(2l+1-k)!(2 m+1-k)!}
\frac{(\frac{(\xi_1+\xi_2)}{2})^{2l+1-k}}{(\frac{2l+1-k}{2})!}\times\nonumber\\
&&\sqrt{(2l+1-k)!}
\frac{(\frac{(\xi_1+\xi_2)}{2})^{2m+1-k}}{(\frac{2m+1-k}{2})!}
\sqrt{(2m+1-k)!} \delta_{2i,2m+1-k} \delta_{2i,2l+1-k}.
\end{eqnarray}

Counting photons is more demanding experimentally than not resolving photon numbers, but photon chopping \cite{chopping} as realized by a  time resolved multiport splitter \cite{choppingexp} may be possible, and could be used at least for small photon numbers, implying that the weakly squeezed states could be compared. In the following we will limit our considerations to coherent states.

\subsection{Several coherent states}

The beam splitter method of comparing two coherent states can easily be generalized to more than two states. For this we need to use a balanced multiport (see Fig. 3),
effecting the transform
\begin{figure}
\includegraphics[width=5.cm]{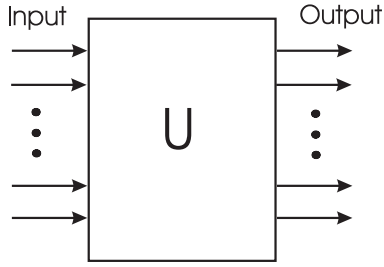}
\caption{The balanced multiport is a passive device distributing an incoming photon with equal probability among all the outputs. The device can be constructed using beam splitters and phase shifters.}
\label{mbs1}
\end{figure}
\begin{equation}
\hat{b}^\dagger_k=\sum_{l=0}^{N-1}u_{kl}\hat{a}^\dagger_l,
\end{equation}
where $\hat{a}^\dagger_l$ are now the creation operators for the $N$ input modes and $\hat{b}^\dagger_k$ are the creation operators for the output modes.
The elements $u_{kl}$ of the transformation matrix of the balanced multiport are given by
\begin{equation}
u_{kl} = \frac{1}{\sqrt{N}}\exp(\frac{2 \pi i kl}{N}), \quad k,l = 0,1,...,N-1.
\end{equation}
We may also think of this as a discrete Fourier transform. The $N$ coherent states $|\alpha_0\rangle\otimes|\alpha_1\rangle\otimes...\otimes|\alpha_{N-1}\rangle$ will transform as
\begin{eqnarray}
|\alpha_0\rangle\otimes|\alpha_1\rangle\otimes...\otimes|\alpha_{N-1}\rangle&=&\exp({-\frac{1}{2}\sum_{j=0}^{N-1}|\alpha_j|^2})\exp(\sum_{l=0}^{N-1}\alpha_l\hat{a}^\dagger_l)|0\rangle\nonumber\\
&=& \exp({-\frac{1}{2}\sum_{j=0}^{N-1}|\alpha_j|^2})\exp(\sum_{l=0}^{N-1}\alpha_l\sum_{k=0}^{N-1}u^*_{lk}\hat{b}^\dagger_k)|0\rangle\nonumber\\
&=&\exp({-\frac{1}{2}\sum_{j=0}^{N-1}|\alpha_j|^2})\exp[\sum_{k=0}^{N-1}(\sum_{l=0}^{N-1}\alpha_lu^*_{lk})\hat{b}^\dagger_k]|0\rangle\nonumber\\
&=&|\sum_{l=0}^{N-1}\alpha_l u^*_{l0}\rangle \otimes|\sum_{l=0}^{N-1}\alpha_l u^*_{l1}\rangle\otimes ...
\otimes|\sum_{l=0}^{N-1}\alpha_l u^*_{l,N-1}\rangle.
\end{eqnarray}
If all $\alpha_j$ are equal, only the zeroth output mode will contain any photons. All the other modes will contain vacuum,
since $\sum_{l=0}^{N-1}u^*_{lk}=0$ unless $k=0$. Therefore, if any photons are detected in any of the modes 1 to $N-1$, all the coherent input
states cannot have been identical. The probability of detecting zero photons in the $k$th output mode will be
\begin{equation}
p_k(0) = \exp(-|\sum_{l=0}^{N-1}\alpha_l u^*_{lk}|^2).
\end{equation}
The probability to detect no photons in any of the output modes 1 to $N-1$ is
\begin{equation}
p(0) = p_1(0)p_2(0)\cdot ... \cdot p_{N-1}(0),
\end{equation}
and the success probability will thus be
\begin{eqnarray}
p_{succ} &=& 1-p(0) = 1-\exp(-\sum_{k=1}^{N-1}|\sum_{l=0}^{N-1}\alpha_l u^*_{lk}|^2)\nonumber\\
&=& 1-\exp(-\frac{1}{2N} \sum_{j,l=0}^{N-1}|\alpha_j-\alpha_l|^2),
\label{succNcoherent1}
\end{eqnarray}
where the second line is obtained after a straightforward and not too lengthy calculation. An alternative way of writing
the success probability is
\begin{equation}
p_{succ}
= 1-\left(\prod_{j,l=0}^{N-1}\langle\alpha_j|\alpha_l\rangle\right)^\frac{1}{N} .
\label{succNcoherent2}
\end{equation}
This success probability will again always be larger than that of the optimal universal comparison strategy; this statement is proved in an appendix.

The multiport setup does not only give knowledge of when all the coherent input states are  not identical. Detection of photons in output mode $k$ means that the sum
\begin{equation}
\sum_{l=0}^{N-1}\alpha_l u^*_{lk}
\end{equation}
must be nonzero. Conversely, if $\alpha_l = |\alpha|\sqrt{N}u_{lk}$ for some $k$, then output mode $k$ must be the only mode containing photons. Detection of a photon
in any other mode than mode $k$ indicates that $\alpha_l \neq |\alpha|\sqrt{N}u_{lk}$ for at least one $\alpha_l$.
Setting $k=0$ we again obtain comparison, i.e. a test whether all $\alpha_l$ are nonidentical.

Although the experimental realization of a balanced multiport for
large $N$ is nontrivial, the suggested scheme should be feasible to
implement for small $N$.  A multiport may be built up
using 2$\times$2 beam splitters \cite{paivi}, alternatively a fiber
coupler could be used, at least for $N=3$ \cite{zeilingerpaper}. It
might be also possible to use the time-resolved realization of a
multiport that we already mentioned in connection with photon
counting \cite{choppingexp}.

\subsection{States which will always pass the comparison test}

In the quantum state comparison schemes for two or more coherent states and for squeezed states proposed above, it is guaranteed that if the input states are identical and of the required form, then they will always pass the test. There are, however, also other quantum states which are guaranteed to pass the test, even if they are {\it not} coherent or squeezed states. The same situation arises also for universal quantum comparison \cite{firstcomp, statecomp}. There, in order to compare the states,  we test whether the total state is symmetric or not. Therefore, any total state which is completely symmetric will always pass the universal comparison test, even if the states of the individual quantum states may not be the same.  For two quantum systems, this would be any state of the form $1/\sqrt{2}(|\phi\rangle|\psi\rangle + |\psi\rangle|\phi\rangle)$. This can be easily generalized to universal quantum comparison of more than two states.

When comparing two coherent states, we associate photons detected in output mode $b$ with the input states being different. Any input state, which results in photons exiting only in mode $a$, will always pass the comparison test. It follows, that a complete basis for the input states which will always pass the test, is given by evolving the number states $|n\rangle_a |0\rangle_b$ backwards through the beam splitter. The input states take the form of SU(2) coherent states, which are entangled,
\begin{equation}
\vert\psi_{in}\rangle = \frac{1}{2^{n/2}} \sum\limits^n_{k=0} \sqrt{n\choose k} \vert n-k\rangle_a \vert k\rangle_b.
\end{equation}
These states, as well as linear combinations and mixtures of them, will always pass the comparison test.
For comparison of many coherent states, a complete basis for the states which will always pass the multiport test is likewise given by evolving linear combinations of the number states $|n\rangle_0|0\rangle_1|0\rangle_2...|0\rangle_{N-1}$ back through the multiport. 

As for quantum comparison of squeezed states, the corresponding states are obtained by evolving number states $|m\rangle_a|n\rangle_b$, where $m$ and $n$ are even, back through the beam splitter. A basis for the input states which always would pass the squeezed state comparison is given by the states
\begin{equation}
\vert\psi_{in}\rangle = \frac{1}{\sqrt{2^m m!}\sqrt{2^n n!}}
\sum_{k,l} {m\choose k} {n\choose l} (-1)^l \sqrt{(m+n-k-l)!}
\sqrt{(k+l)!} \vert m+n-k-l\rangle_a\vert k+l\rangle_b .
\end{equation}
These states are again entangled states of the input modes. Linear combinations of, and statistical mixtures of these states (and statistical mixtures of linear combinations of these states) will always pass the comparison test for squeezed states.

Next, we analyze two simple quantum cryptographic protocols where quantum state comparison of coherent states is needed. The first scheme, denoted as a quantum ``lock and key" scheme, is based on the seminal work of S. Wiesner \cite{wiesner}, which sparked the field of quantum cryptography. The second example we consider is based on ideas for public-key cryptography. More precisely, we introduce a protocol to distribute and test quantum public keys.

\section{A quantum ``lock and key" scheme}

In his original proposal, S. Wiesner showed how to use quantum-mechanical systems in order to create a secret key that is impossible to counterfeit, but which can be validated by means of a lock \cite{wiesner}. The main idea behind this scheme is to use, as a secret key, a sequence of $M$ 
quantum systems, each  one prepared in a state that is selected, randomly and independently,
within a given set of $N$ non-orthogonal quantum states. Here we will consider a set of non-orthogonal states composed only of coherent states $|\alpha_j\rangle$, as we have in mind the experimental realization described in the previous section, {\it i.e.},
\begin{equation}
|\psi_{key}\rangle = |\alpha_{1}\rangle\otimes|\alpha_{2}\rangle\otimes ... \otimes|\alpha_{M}\rangle.
\end{equation}

Each quantum key $|\psi_{key}\rangle$ is associated with a unique quantum lock state $|\psi_{lock}\rangle$, composed of an identical string of coherent states, {\it i.e.}, $|\psi_{lock}\rangle=|\psi_{key}\rangle$. In order to 
check if a given key is valid and opens the lock, one needs to compare the key string with the lock string. All the key states must match the corresponding lock states, or more precisely, none of the key states may be detected as different from the corresponding lock state. As a result, a possible
adversary who is ignorant of the key states has absolutely no way of counterfeiting them faithfully.

Note that this protocol could just as well be implemented by using a classical lock instead of a quantum lock. In this case, the lock would contain a classical record of the actual states in the key string. Now, in order to test whether the key fits in the lock, one can measure each state in the key string, projecting it onto the projectors $|\alpha_j\rangle\langle\alpha_j|$ and $\mathbf{1}-|\alpha_j\rangle\langle\alpha_j|$. This measurement can be effected using  the classical record of the state $|\psi_{key}\rangle$, which is stored in the lock. Quantum comparison of two unknown quantum states would not be needed in this scenario, only measurements performed on single quantum states. The version where the lock contains no classical record, but only the quantum states $|\alpha_j\rangle$, has, however,  the advantage that in this case, it is impossible for an adversary to make new perfect key copies based on the information stored in a lock.

We will assume that all the coherent states $|\alpha_j\rangle$ included in the key have the same amplitude $|\alpha|$, while the phase of each individual state is chosen randomly and independently as $2\pi k/N$, where $k\in (0,1,2,...,N-1)$, with equal probability for each $k$. Other choices are of course possible. Next, we analyze, in more detail, the security of this ``lock and key" scheme against a possible adversary with unlimited quantum computational power.

\subsection{Forcing the lock open without a key}

An adversary who does not have a key, and who does not know the phase of each individual $\alpha_j$, could still try to open the lock. We will assume that the information about the amplitude
$|\alpha|$ is public. The adversary is not limited to using coherent states in order to try to counterfeit a key, but can prepare any general quantum state, where the states of the individual positions might be entangled. Note, however, that since the phases of the coherent states in each lock position are random and uncorrelated, and the comparison test is performed for each position of the lock string individually, he or she cannot get any advantage from using entangled states. Assuming that the states in the individual key positions are not entangled, then, for each position in the key, the adversary can prepare a general state 
$\int_{-\infty}^\infty d^2\beta P(\beta)|\beta\rangle\langle\beta |$. Here $\int_{-\infty}^{\infty} d^2\beta = \int_{-\infty}^{\infty} \int_{-\infty}^{\infty}  d\beta_rd\beta_i$, with $\beta_r = \text{Re}\beta$ and $\beta_i = \text{Im}\beta$, and the adversary is choosing $P(\beta)$ so that the probability to pass the comparison test is as high as possible. $P(\beta)$ is the $P$-function of the state, and any state can be written in this way with a suitably chosen (albeit sometimes highly singular) $P$-function.  The probability for the
false key state to pass the comparison test with the lock state in one position, which is $|\alpha_j\rangle = ||\alpha| e^{i\theta}\rangle$, is, on average,
\begin{equation}
p_{pass} = 1-p_{succ} =  \frac{1}{2\pi}\int_0^{2 \pi} d\theta \int_{-\infty}^\infty d^2\beta P(\beta) \exp(-\frac{1}{2}\left||\alpha| e^{i\theta} -\beta\right|^2),
\end{equation}
where we integrate over $\theta$, since the phase $\theta$ is chosen randomly with a uniform distribution, and the adversary does not know the phase, only the amplitude, of $\alpha_j$. For simplicity, the number $N$ of possible phase angles is infinite in the expression above, but one could also calculate $p_{pass}$ for a specific $N$. In a real protocol, $N$ should, in any case, be large. The adversary wants to maximize the probability $p_{pass}$. Writing $\beta = |\beta |e^{i\theta_\beta}$, 
and  assuming that we can switch the order of integration, we obtain
\begin{eqnarray}
p_{pass} &=& \frac{1}{2\pi}\int_0^{2 \pi} d\theta \int_{-\infty}^\infty d^2\beta {P}(\beta) \exp(-\frac{1}{2}\left||\alpha| e^{i\theta} -|\beta|e^{i\theta_\beta}\right|^2)\nonumber \\
&=& \frac{1}{2\pi}\int_{-\infty}^\infty d^2\beta {P}(\beta) \int_0^{2 \pi} d{\theta}\exp[-\frac{1}{2}(|\alpha|^2+|\beta|^2-2|\alpha\beta|\cos({\theta-\theta_\beta)})]\nonumber\\
&=& \int_{-\infty}^\infty d^2\beta P(\beta) \exp[-\frac{1}{2}(|\alpha|^2+|\beta|^2)]I_0(|\alpha\beta|),
\end{eqnarray}
where 
the function $I_0(|\alpha\beta|) = \frac{1}{2\pi}\int_0^{2 \pi} d{\theta}\exp(|\alpha\beta|\cos\theta) = \frac{1}{2\pi}\int_0^{2 \pi} d{\theta}\exp[|\alpha\beta|\cos(\theta-\theta_\beta)]$ is a modified Bessel function of the first kind. 
It turns out that, for $\alpha \leq \sqrt{2}$, $p_{pass}$ is maximized if we choose $P(\beta) = \delta(0)$, that is, the best false key state is a vacuum state and the maximum probability to pass the comparison test is given by
\begin{equation}
p_{pass} = \exp(-\frac{1}{2}|\alpha|^2).
\end{equation}
For $\alpha \geq \sqrt{2}$, the maximum probability to pass occurs if the adversary chooses $|\beta|$ closer to $|\alpha|$, but still with $|\beta|<|\alpha|$. For large values of $|\alpha\beta|$, $I_0(|\alpha\beta|)\sim e^{|\alpha\beta|}/\sqrt{2\pi |\alpha\beta|}$, and therefore
\begin{equation}
\exp[-\frac{1}{2}(|\alpha|^2+|\beta|^2)]I_0(|\alpha\beta|) \sim \frac{1}{\sqrt{2\pi |\alpha\beta|}}\exp[-\frac{1}{2}(|\alpha|-|\beta|)^2].
\end{equation}
Therefore, for large $\alpha$, the adversary should choose $|\beta|\approx|\alpha|$ to maximise the probability to pass the comparison test. This probability will decrease as a function of $\alpha$ approximately as
\begin{equation}
p_{pass} \sim \frac{1}{\sqrt{2\pi}|\alpha|}.
\end{equation}
For a key string containing $M$ coherent states, the probability for a false key state to pass the comparison test for all $M$ positions is $p_{pass}^M$, so that the probability decreases exponentially with $M$. As long as $\alpha$ is not too small, we find that the probability of the adversary successfully cheating is severely restricted. In addition, if the total average number of key and lock photons at the output is measured (or the number of key photons is measured directly at the input), then any cheating strategy where a false key state contains the wrong average number of photons would be discovered.  

\subsection{Obtaining information about a key}

Let us now suppose that an adversary has access to one valid copy of the key, and that he or she tries to extract information from it. Of course, it is clear that once an adversary 
has a valid key, then this key can always be used to open the lock. But if a key copy is missing, this may be noticed by the rightful owner of the key. Obtaining a full classical description of the quantum state
of the key, on the other hand,  would allow the adversary to prepare as many valid keys for a given lock as he/she wishes. In particular, the adversary could make one copy for returning to the rightful owner, so that it is perhaps not noticed that a key copy has been stolen, as well as extra ``illegal" key copies. We will now show that the information that can be obtained by measurements on one or more copies of a key is limited.

The maximum information that the adversary can obtain by measurements on one single copy of the key string, called accessible information and denoted as $I_{acc}$, is limited by the Holevo quantity $\chi(\rho_{key})$. Here $\rho_{key} =  \sum_n p_n \rho_n $ is the state of the key string according to the information available to the adversary before the measurement; in other words, $\rho_n$ are the possible states of the key string, and $p_n$ their respective probabilities. The possible key states $|\alpha_j\rangle$ in each position are given by $||\alpha| \exp(2 \pi i k/N)\rangle$, where $k$ takes the values $0,1,2, ...,N-1$. If there are $M$ positions in the key string, then, as far as the adversary knows, there are $N^M$ possible pure states $\rho_n$, all equiprobable, with $p_n = 1/(N^M)$, that the total key string could have.
The accessible information about which of these $N^M$ states the key state actually is, is bounded according to
\begin{equation}
I_{acc}\leq \chi(\rho_{key}) = S(\rho_{key}) -\sum_n p_n S(\rho_n),
\end{equation}
where $S(\rho_{key})=-{\rm Tr}(\rho_{key}\log_2\rho_{key})$ is the von Neumann entropy of $\rho_{key}$. The quantity $\sum_n p_n S(\rho_n)$ is always positive or
zero. When the different possible states $\rho_n$ are pure, as in our case, it is
zero.

\begin{figure}
\includegraphics[width=7cm]{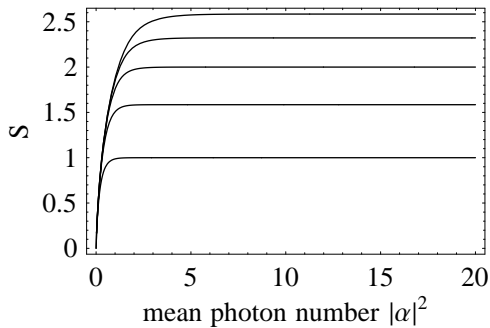}
\caption{The von Neumann entropy as a function of the mean photon
number $\vert\alpha\vert ^2$ for $N=2,3,4,5,6$. The asymptotic 
value of the entropy increases with the number of states $N$.} \label{Fig4}
\end{figure}

As the $M$ coherent states in different positions in the key string
are completely uncorrelated, the accessible information of the whole
key string is bounded by $M$ times the accessible information for
each position in the key string. Let us therefore look at the state
in a single key position. Since the adversary does not know the
phase of the coherent key state $|\alpha_j\rangle$, in this position, the density matrix according to the information available about the state prior to the measurement is
\begin{equation}
\label{guessrhofinite}
\rho_{single} = \frac{1}{N}\sum_{k=0}^{N-1}|\alpha e^{i k 2\pi/N}\rangle\langle\alpha e^{i k 2\pi/N}|.
\end{equation}
For this state, the von Neumann entropy, which limits the accessible information since the different possible states are pure, can be found to be
\begin{equation}
S(\rho_{single}) = \sum_{m=0}^{N-1}\frac{1}{N K_m^2} \log_2(N K_m^2),
\end{equation}
where
\begin{equation}
K_m^{-2} = \sum_{k=0}^{N-1} \exp\{-|\alpha|^2[1-\exp(ik2\pi/N)]+imk2\pi/N\}.
\end{equation}
When $|\alpha|=0$, $\rho_{single}=|0\rangle\langle 0 |$. As there is
only one possible state, the information stored in the key state is
zero in this case. The von Neumann entropy and the accessible
information for the key state are also zero. For a useful key
scheme, we need to choose $|\alpha|$ larger than zero, but not too
large, for a given number of states $N$. For instance, when the amplitude $|\alpha|$ goes to infinity,
the accessible information for each key position approaches $\log_2
N$, which is the information that can be obtained from $N$
distinguishable states, or from $N$ classical states. This reflects
the fact that, for larger $|\alpha|$, the $N$ possible key states
become more distinguishable. A coherent state with a larger
amplitude becomes more ``classical". In Fig. \ref{Fig4}, the von Neumann
entropy, which bounds the accessible information, is plotted as a
function of $|\alpha|^2$ for some values of $N$.

We should choose the amplitude $|\alpha|$ small enough, and $N$ large
enough, for the accessible information not to be too large compared to $\log_2 N$, but
$|\alpha|$ large enough 
for the probability to detect a difference in the key and lock states to be sufficiently
large. In particular, as we have seen that the best false key state is a vacuum state, we have to adjust $|\alpha|$ so that there is a reasonable probability to detect this cheating strategy. 
As we saw previously, the probability to detect a difference
in key and lock can also always be increased by increasing the
length $M$ of the key string.

When $N$ goes to infinity, the state in equation
(\ref{guessrhofinite}) becomes a phase-randomized state, which is
diagonal in the number state basis and can be written as
\begin{equation}
\rho_{single}^\infty = e^{-|\alpha|^2}\sum_{k=0}^\infty \frac{|\alpha|^{2k}}{k!} |k\rangle\langle k|.
\end{equation}
The von Neumann entropy for this state is
\begin{equation}
S(\rho_{single}^\infty) = |\alpha|^2 - e^{-|\alpha|^2} \sum_{k=0}^\infty \frac{|\alpha|^{2k}}{k!}\log_2\left(\frac{|\alpha|^{2k}}{k!}\right).
\end{equation}
As before, this quantity also bounds the accessible information.
We can obtain an approximation for this expression using the Stirling formula
for the factorial. The result takes the rather simple form
\begin{equation}
S(\rho_{single}^\infty) \approx \log_2 (2\pi e \vert\alpha\vert^2) .
\end{equation}
The entropy increases in  a logarithmic way with the coherent state amplitude $|\alpha|$.

Till now, we have considered a ``lock and key" scheme where only one
single copy of each key may exist. However, it is possible to design
a protocol which uses as many copies of the key as we like. The
security of the scheme will necessarily decrease with the number of
key copies. An adversary wanting to fabricate illegal key copies
could get hold of all the keys in circulation, and using these, will
be able to fabricate a better false key than if just one or very few
key copies are in circulation. However, the information an adversary
can obtain per key copy is still limited by the Holevo bound. If it
is possible to obtain at most $K$ bits of information about the state in one position of the key
 when one copy is available, then at most $TK$ bits can be
obtained if $T$ copies are available. In this last case, we need to
guarantee that $\log_2 N\gg TK$.

Finally, let us briefly mention that the adversary might try to make a copy of the single existing key. For this, he would need to make a clone of each individual coherent state. 
This is only possible with a certain degree of fidelity, as making perfect copies is forbidden by the no-cloning theorem. We should, however, bear in mind that we do not deal with completely unknown states but with a know class of states -- large enough amplitude coherent states can be
copied almost perfectly. Cheating by this method can, however, be prevented by choosing a long enough string of states, or by choosing $N$ large enough.

\section{Quantum public key distribution}

Public key cryptography requires two keys---the public key and the
private key, which form a key pair. The sender, usually called
Alice, generates the key pair, makes the public key public and keeps
her private key in a secret place to ensure its private possession.
The key generation algorithm is designed in such a way that anyone
having a public key can, for instance, use it in order to encrypt a
message for Alice (public key encryption schemes), or certify that a
message originates from Alice (digital signature schemes). However,
only Alice can decrypt or sign a message using her private key
\cite{Schneier}.

Unfortunately, the security of classical public key cryptography
rests on unproven assumptions related to the intractability of
certain difficult mathematical problems. The key generation
algorithm utilizes so-called one-way functions to guarantee
that the public keys do not reveal information about the private
key. This kind of mathematical functions are easy to evaluate in
one direction, but their inverse is very difficult to compute
\cite{diffie,rivest}. However, these computational assumptions may
be defeated by exhaustive computer analysis, or by the discovery
of better algorithms for solving the problems on which they are
based. If a quantum computer is ever built, many classical public
key cryptosystems in use today will become unsafe, leading also to
a retroactive security break \cite{shor}.

Quantum-mechanical systems 
can be used to create one-way functions which are provably secure from
an information-theoretic point of view. For instance, one can obtain
a quantum one-way function by defining a quantum map ${\cal F}:\
k\in\{0,1\}^n\rightarrow\ |\psi_k\rangle$, whose input is a
classical $n$-bit string $k$, and whose output is a quantum state
$|\psi_k\rangle$ \cite{digital}. As in the previous section, we
will here consider that $|\psi_k\rangle$ is of the form
$|\psi_k\rangle=|\alpha_{1}^k\rangle\otimes|\alpha_{2}^k\rangle\otimes
... \otimes|\alpha_{M}^k\rangle$, where the state of each coherent
state $|\alpha_j^k\rangle$ belongs to a given public set of $N$
possible coherent states. In this case, we have that $n=M\log_2 N$.
The impossibility of inverting the function ${\cal F}$ can be
guaranteed by means of the Holevo bound, which limits the amount of
classical information that can be extracted from a quantum state. In
particular, and assuming that there are $T$ copies available of each
public key $|\psi_k\rangle$, we find that an adversary can obtain,
at most, $TS(\rho_{public})$ bits of information by measuring all
the copies of the public key, where
$\rho_{public}=1/(N^M)\sum_{k}|\psi_k\rangle\langle\psi_k|$
represents the state of the public key according to the information
available about it before the measurement, and $S$ is the von
Neumann entropy. That is, if we assure that $n=M\log_2 N\gg
TS(\rho_{public})$, then the probability of successfully guessing
the classical private key $k$, given {\it all} the public keys,
remains small. Note that in the case of quantum public keys, this
means that only a limited number $T$ of them can be in circulation
to guarantee unconditional security.

Next, we present two possible schemes that use balanced multiports
to securely distribute quantum public keys. The first scheme assumes
the availability of a trusted key distribution center which has
authenticated links \cite{barnum} to all the participants. In the
second scheme, we consider the scenario where all the recipients
obtain their public keys directly from Alice via an authenticated
quantum channel, and no trusted key distribution center is
available. We study the security of both schemes against two
scenarios of cheating, motivated from the key distribution
phase which is needed in the quantum digital signature scheme
introduced in Ref.~\cite{digital}. For simplicity, in the security
analysis we will consider the case where there are only two
recipients, called Bob and Charlie. The extension to a higher
number of recipients is straightforward. In the first cheating
scenario, only Alice is dishonest; her objective, once the public
key distribution phase is completed, is to get Bob and Charlie to
disagree about the validity of the private key when this key is
revealed. In a digital signature scheme, this case corresponds to
Alice trying to repudiate the signature of a message with her
private key. In the second cheating scenario, Alice and at least
Bob are honest, while Charlie can be dishonest. The goal of
Charlie is to make Bob accept as valid a false public key that
does not come from Alice, but comes from Charlie. This corresponds 
to the standard forging scenario. Note that
Charlie could always prevent Bob from receiving any public key
coming from Alice just by cutting the line, but we do not consider
this to be a success for the cheaters.

\subsection{Public key distribution with trusted center}

The goal is to generate and distribute $T$ copies of the quantum
public key $|\psi_k\rangle$ selected by Alice. One straightforward
solution in order to do this is to assume the existence of a trusted
key distribution center composed by $M$ balanced multiports with $T$
inputs each. Alice prepares and sends to the key distribution center
the quantum state
$|\psi_k^T\rangle=|\sqrt{T}\alpha_{1}^k\rangle\otimes|\sqrt{T}\alpha_{2}^k\rangle\otimes
... \otimes|\sqrt{T}\alpha_{M}^k\rangle$ as a starting point for
generating the public keys. Once this state is received by the trusted
center, each coherent state $|\sqrt{T}\alpha_{j}^k\rangle$, with
$j=1...M$, is used as one input for the $j$th balanced multiport,
while the remaining $T-1$ inputs of each multiport contain vacuum.
As a result, the output state of the $j$th multiport is given by
$|\alpha_{j}^k\rangle^{\otimes{}T}$, {\it i.e.}, it contains $T$
copies of the coherent state $|\alpha_{j}^k\rangle$. Combining all
the output states of the $M$ multiports in the trusted center one
obtains the state $|\psi_k\rangle^{\otimes{}T}$. To conclude, the
trusted center sends each receiver one copy of the public key
$|\psi_k\rangle$ through an authenticated quantum channel.

Let us now analyze the security of this public key distribution scheme according to the cheating strategies introduced above. A reader who is not interested in the security proof may go directly to Sec. \ref{nocenter}.
Since each public key $|\psi_k\rangle$ is sent to each receiver
via an authenticated quantum channel established with the trusted
center, it is clear that a dishonest Charlie cannot make Bob
receive a false public key. We need, therefore, to consider only
the case where Alice is dishonest. That is, we have to evaluate
the probability of Bob and Charlie to disagree about the validity
of the private key after using the public key distribution scheme
introduced above. Here we consider the case where the private key
is made public in a later step of the particular quantum
cryptographic protocol that uses the public keys obtained from the
trusted center \cite{digital}. Note that we are only interested in
the security of the quantum public key distribution protocol.

After Alice announces her private key $k'$ (or a function of it), Bob and Charlie can compute the function $\cal F$ and obtain a classical description of the corresponding public key $|\psi_{k'}\rangle$. We use a different index $k'$, since Alice, or somebody else, could try to distribute a private key that does not match the previously distributed public key. Now, in order to evaluate whether $k'$ is correct and originates from Alice, they can test whether the state $|\psi_{k'}\rangle$ is equal to the public keys obtained previously from the trusted center. This test can be done, for instance, by projecting each single position $j$ in the string of states of the public key onto the projectors
$|\alpha_j^{k'}\rangle\langle\alpha_j^{k'}|$ and
$\mathbf{1}-|\alpha_j^{k'}\rangle\langle\alpha_j^{k'}|$ coming from the knowledge of $|\psi_{k'}\rangle$. Then each recipient can count the number of positions $j$ where the measurement test provides an incorrect result, {\it i.e.}, a result associated with the projector
$\mathbf{1}-|\alpha_j^{k'}\rangle\langle\alpha_j^{k'}|$. We denote the number of incorrect results by $e$.

When all the parties are honest, all the recipients obtain $e=0$. If Alice is dishonest, then the quantum public key distribution protocol presented above cannot prevent a situation where one recipient obtains $e=0$, while others obtain $e>0$ with high probability. For instance, Alice could send to the trusted center a quantum state $|\psi_k^T\rangle$, which differs from $|\psi_{k'}\rangle$ in only one position. This position could contain a coherent state $|\sqrt{T}\beta\rangle$ satisfying $|\langle\beta|\alpha^{k'}\rangle|^2=1/2$, where the coherent state $|\alpha^{k'}\rangle$ denotes the state of $|\psi_{k'}\rangle$ in that position. For this simple scenario, we find that Bob and Charlie will obtain, respectively, $e=0$ and $e=1$ (or vice versa) with probability $1/2$. Moreover, note that a dishonest Alice is not restricted to use coherent states in order to prepare $|\psi_k^T\rangle$, but she can use any general quantum state. What this public key distribution protocol can guarantee with high probability, however, is that if one receiver
obtains $e=0$, then no other receiver will obtain $e>sM$ for $s$ or $M$ sufficiently large. Here $s\in[0,1]$ represents a security parameter of the key distribution protocol. This result can be used in a cryptographic protocol, which uses the public keys coming from the trusted center, to guarantee the following \cite{digital}. If no errors are found, {\it i.e.}, $e=0$, the recipient ({\it e.g.} Bob) can conclude that $k'$ is correct, and he can be sure (with high probability) that any other recipient ({\it e.g.} Charlie) will also conclude that $k'$ is correct. If $0<e<sM$, Bob can, also in this case, conclude that the key is correct, but now he cannot be sure that a second recipient (Charlie) will not conclude that $k'$ is incorrect. Finally, if $e\geq{}sM$, Bob can consider the private key to be incorrect, and that Charlie
would either also consider it to be incorrect, or at least Charlie would know that Bob may conclude that $k'$ is incorrect.

Next we obtain an upper bound on the probability of Alice to cheat.
In order to do that, let us first consider the following situation.
Imagine that the trusted center knows the private key $k'$ that
Alice is going to declare later on, and, instead of distributing to
Bob and Charlie the two quantum public keys coming from the
multiports, he sends them directly the classical results obtained
from measuring each of these two quantum public keys accordingly to
the string of states contained in $|\psi_{k'}\rangle$. That is, he
sends Bob and Charlie the classical results of projecting each
position of the public keys onto the the projectors
$|\alpha_j^{k'}\rangle\langle\alpha_j^{k'}|$ and
$\mathbf{1}-|\alpha_j^{k'}\rangle\langle\alpha_j^{k'}|$. Moreover,
for each position $j$ in the key string $|\psi_{k'}\rangle$, the results
obtained from the measurements on the two public keys are
distributed to Bob and Charlie at random. 

Note, now, that the fact that the trusted
center, instead of Bob and Charlie, measures the public keys, 
does not modify the measurement statistics that Bob and Charlie 
would obtain in the original scenario, once $k'$ is known and they
perform their measurements according to $|\psi_{k'}\rangle$.
Moreover, the random distribution of the classical results 
is guaranteed by the intrinsic random character of the multiport 
used by the center to distribute the states to Bob and Charlie. 
Alice makes Bob and Charlie disagree if one of them obtains $e=0$ (in absence of noise) and the other obtains $e\geq{}sM$. This means that the probability of Alice to cheat in this particular situation, $p_{cheat}$, is maximized if, in total, the trusted center finds only $sM$ errors in both public keys, and
he sends all the errors to Bob or to Charlie. We obtain, therefore, 
\begin{equation}\label{Prob_Alice_cheat}
p_{cheat}\leq{}\big(\frac{1}{2}\big)^{sM-1}.
\end{equation}
This upper bound also represents an upper bound on the probability of Alice to cheat in general. 

\subsection{Public key distribution without trusted center}
\label{nocenter}

Let us now analyze the scenario where no trusted center is available. 
In this case, Alice sends one copy of the public key $|\psi_k\rangle$ 
directly to each recipient, via an authenticated quantum
channel. Then, in order to prevent Alice from cheating, all the
recipients need to collaborate to verify that all the public keys
sent by Alice are equal. In order to do this, they use a distributed
comparison test, which can be divided in two phases. Essentially, each recipient compares his or her public key copy with all the other recipients' copies, and, if Alice has sent different public key copies to different recipients, this will be detected. Neither can any of the recipients sabotage the public key copy of another recipient. This would also be detected in the comparison test. 

\begin{figure}
\includegraphics[width=8.cm]{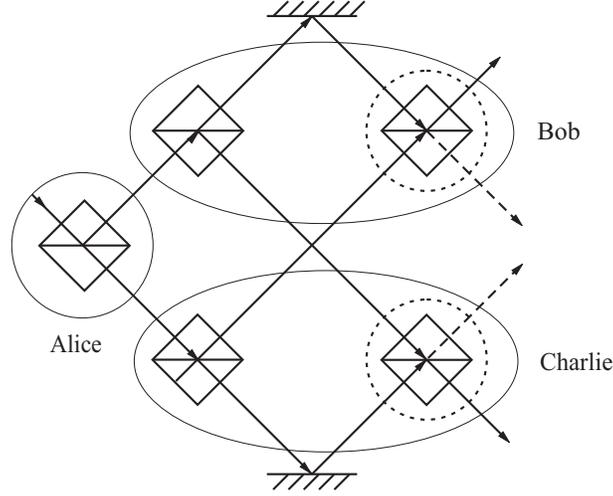}
\caption{The setup for public key distribution without a trusted center, when there are two recipients, Bob and Charlie. Alice sends Bob and Charlie one copy each of her public key. In the picture, she uses a beam splitter to do this. Bob and Charlie then split their key copies in two using beam splitters, and exchange ``key halves" with each other. They then perform comparison tests, indicated by the dashed circles, on their own half key copy and the one they received from the other recipient. If all parties are honest (and their detectors perfect), the output ports with the dashed arrows should only contain vacuum.}
\label{keydist}
\end{figure}

The protocol requires that each recipient has $2M$ balanced multiports with $T$ inputs each.
In the first phase, the first $M$ balanced multiports are used to
split the quantum public key sent by Alice. The case for two recipients is shown in Fig. \ref{keydist}.
In concrete, each
coherent state $|\alpha_{j}^k\rangle$ in $|\psi_k\rangle$, with
$j=1,..., M$, is used as one input for the $j$th balanced multiport,
while the remaining $T-1$ inputs of each multiport contain vacuum.
The output state of this multiport is given by
$|(1/\sqrt{T})\alpha_{j}^k\rangle^{\otimes{}T}$. That is, it
contains $T$ copies of the coherent state
$|(1/\sqrt{T})\alpha_{j}^k\rangle$. Now, each recipient keeps for
himself one copy of $|(1/\sqrt{T})\alpha_{j}^k\rangle$, and
distributes the remaining $T-1$ copies of it to the other $T-1$
recipients via an authenticated quantum channel. The second phase
includes a quantum state comparison test using the second set of
$M$ multiports. The $j$th multiport in this second set receives as
input the coherent state $|(1/\sqrt{T})\alpha_{j}^k\rangle$ kept
by the recipient after the first phase, together with the
corresponding $T-1$ ``copies" of it  obtained from the other $T-1$
recipients. Note that, if all the parties are honest, the zeroth
output mode of this multiport will contain the state
$|\alpha_{j}^k\rangle$, while all the other modes will contain
vacuum. That is, combining all the output states of these $M$
multiports, each recipient can recover Alice's quantum public key
$|\psi_k\rangle$ perfectly. The non-demolition character of the quantum comparison procedure (meaning that it does not alter or destroy the compared states if all parties are honest, so that the compared coherent states are identical to start with) is seen to be vital for the protocol to work.

Next, we study the situation when Alice is dishonest. A reader who is not interested in the security proof can go directly to the Conclusions. In
principle, instead of preparing $T$ copies of $|\psi_k\rangle$ and
distributing them among the legitimate recipients, Alice can prepare
any general quantum state, including entangled states. 
However, it turns out that, for each position $j$ in the public
key strings sent by Alice, the states that the recipients obtain
as output of the zeroth mode of the $j$th multiport used for state
comparison are completely symmetric under permutation. This means
that, although Alice could in principle prepare states that make
the parties disagree about the validity of the private key $k'$,
she cannot control which of the recipients receives the valid
results. Once Alice declares the value of $k'$ and the recipients
project their public keys, coming from the output of the zeroth
mode of the $M$ multiports, onto the projectors
$|\alpha_j^{k'}\rangle\langle\alpha_j^{k'}|$ and
$\mathbf{1}-|\alpha_j^{k'}\rangle\langle\alpha_j^{k'}|$, the
errors are distributed at random between all the recipients without
Alice being able to control this. We can use, therefore, the same
argumentation as in the previous section, to obtain that the
probability of Alice to cheat in this scenario also satisfies
Eq.~(\ref{Prob_Alice_cheat}).

Finally, we need to consider the situation where Alice and
Bob are honest, but Charlie can be dishonest. The goal of Charlie is
to make Bob accept a false public key that does not match Alice's
private key. Note that in this key distribution protocol, a dishonest
Charlie can try to influence Bob's public key by means of the
quantum states that he needs to send to Bob for the comparison test. In
order to make Bob reject Alice's private key, Charlie needs to send
him quantum states that can produce at least $e\geq{}sM$ errors in
Bob's results. However, for sufficiently large $s$ or $M$, this
situation can also be detected by Bob in the comparison test.
Whenever Charlie sends Bob a state different from the one coming
from Alice, Bob can detect this fact by finding photons not only on
the zeroth output mode of the corresponding multiports used for
comparison.

\section{Conclusions}

We have analyzed quantum state comparison for the case when one has prior knowledge about the class of states from which the states to be compared are chosen. We chose to look at comparison of coherent states, and have shown that, for large coherent state amplitudes, the probability to detect that the two coherent states are different, when they are indeed different, approaches one (certainty). In contrast to this, the success probability for a universal comparison strategy never exceeds 1/2.  A universal strategy has to be used when no prior information about the quantum states is available. In addition to the high success probability, the quantum comparison strategy for coherent states has a non-demolition character - it  does not destroy the compared quantum states, if they are indeed equal coherent states. In this case, one can recover the original coherent states unaltered. If the compared coherent states are unequal to start with, then they will be altered by the procedure.

Following this, coherent state comparison was used to develop two examples of applications --- a ``lock and key" scheme and a public key distribution scheme. For both these applications, the non-demolition character of the quantum comparison procedure is vital. We believe that both examples are not only conceptually simple, but also of some practical importance due to their experimental accessibility.

\section*{Acknowledgements}

IJ gratefully acknowledges financial support by GA\v CR 20/04/2101, LC 060001 MSMT and the 6th FP Quele. MC gratefully acknowledges financial support from the DFG under the Emmy Noether programme, and the European Commission (Integrated Project SECOQC). EA gratefully acknowledges the Royal Society of London for financial support.
EA and IJ thank N. L\"utkenhaus for hospitality during their stays in Erlangen, and all three authors thank him for the many fruitful discussions on the topics included in this paper.

\begin{appendix}
\section{Success probability for quantum comparison of $N$ coherent states}
In this appendix, we will prove that the quantum comparison strategy, which is tailored for coherent states, always has a larger success probability than the universal quantum comparison strategy, when comparing $N$ given coherent states $|\alpha_0\rangle, |\alpha_1\rangle , ..., |\alpha_{N-1}\rangle$.
The universal comparison strategy is a projection onto the totally symmetric, and onto the asymmetric
subspaces. If the total state of the quantum systems is found to be asymmetric, then the states of the individual systems
cannot all have been the same. The success probability of the coherent state strategy is given by Eqns. (\ref{succNcoherent1}) and (\ref{succNcoherent2}). 
The success probability of the optimal universal strategy will be \cite{statecomp, tonycomp}
\begin{equation}
p_{asymm} = 1-p_{symm} = 1-\langle\alpha_0|\langle\alpha_1|...\langle\alpha_{N-1}|P_{symm}|\alpha_0\rangle|\alpha_1\rangle ...|\alpha_{N-1}\rangle,
\end{equation}
where $P_{symm}$ is the projector onto the symmetric subspace. We have that
\begin{equation}
p_{symm} = \frac{1}{N!}\langle\alpha_0|\langle\alpha_1|...\langle\alpha_{N-1}|\sum_{perm} |\alpha_{i_0}\rangle|\alpha_{i_1}\rangle ...|\alpha_{i_{N-1}}\rangle,
\label{Nstates}
\end{equation}
where the sum should be taken over all $N!$ permutations of the indices in the kets, so that $(i_0,i_1,i_2,...,i_{N-1})$ is a permutation
of $(0,1,2,...,N-1)$. As an example, for $N=3$,
\begin{eqnarray}
p_{symm} &=& \frac{1}{3!}(1 + |\langle\alpha_0|\alpha_1\rangle|^2 + |\langle\alpha_1|\alpha_2\rangle|^2 +|\langle\alpha_2|\alpha_0\rangle|^2\nonumber\\
&+&\langle\alpha_0|\alpha_1\rangle\langle\alpha_1|\alpha_2\rangle\langle\alpha_2|\alpha_0\rangle+\langle\alpha_0|\alpha_2\rangle\langle\alpha_1|\alpha_0\rangle\langle\alpha_2|\alpha_1\rangle) .
\label{threestates}
\end{eqnarray}
To prove that the multiport strategy for coherent states always will have a greater success probability than the universal strategy, we will use the fact that for a collection of $N$ numbers, their geometric mean, defined as the $N$th root of their product, is always smaller than their arithmetic mean. Starting with the case of three coherent states, $p_{symm}$ in equation (\ref{threestates}) can be viewed as the
arithmetic mean of the six terms in the parenthesis. The geometric mean of these six numbers is
\begin{equation}
\left(|\langle\alpha_0|\alpha_1\rangle|^4|\langle\alpha_1|\alpha_2\rangle|^4|\langle\alpha_2|\alpha_0\rangle|^4\right)^{1/6} = \left(\prod_{j,l=0}^{2}|\langle\alpha_j|\alpha_l\rangle|\right)^\frac{1}{3} ,
\end{equation}
which is the probability that the coherent-state multiport scheme will fail for three coherent states. The coherent-state multiport scheme therefore has a larger probability to succeed than the universal quantum comparison strategy. For general $N$, the proof is similar. The quantity $p_{symm}$ in equation (\ref{Nstates}) is viewed as the arithmetic mean of $N!$ numbers. To calculate the geometric mean of these numbers, we need their product. In this product, the factor $\langle\alpha_j|\alpha_l\rangle$ will occur $(N-1)!$ times, since if we choose to pair $j$ with $l$, there are $(N-1)!$ ways to choose the rest of the index pairs. Therefore the geometric mean of the $N!$ numbers is
\begin{equation}
\left(\prod_{j,l=0}^{N-1}\langle\alpha_j|\alpha_l\rangle^{(N-1)!}\right)^{1/N!} = \left(\prod_{j,l=0}^{N-1}\langle\alpha_j|\alpha_l\rangle\right)^{1/N}=1-p_{succ} \leq p_{symm},
\end{equation}
which means that the multiport comparison strategy for coherent states has a smaller probability to fail than the universal quantum comparison strategy --- in other words, it will always do better.

\end{appendix}

\end{document}